\documentclass[10pt]{article}
\usepackage{amsfonts}
\title{Another Critique of the Replica Trick} 
\author{Martin R. Zirnbauer\footnote{Current address:
  Department of Mathematics, Fine Hall, Princeton University, USA.
  Permanent address: Institut f\"ur Theoretische Physik, Universit\"at
  zu K\"oln, Z\"ulpicher Str. 77, 50937 K\"oln, Germany.  
  Email: zirn@thp.Uni-Koeln.DE}}
\begin{document}
\date{March 20, 1999}
\maketitle
\begin{abstract}
Kamenev and Mezard, and Yurkevich and Lerner, have recently shown how
to reproduce the large-frequency asymptotics of the energy level
correlations for disordered electron systems, by doing perturbation
theory around the saddles of the compact nonlinear $\sigma$ model
derived from fermionic replicas.  We present a critical review of their
procedure and argue that its validity is limited to the perturbative
regime of large frequency.  The miraculous exactness of the
saddle-point answer for $\beta = 2$ (unitary symmetry) in the
universal limit, is shown to be a special feature due to the
Duistermaat-Heckman theorem.
\end{abstract}

\section{Introduction}

Suppose we are given a disordered statistical mechanical system with
free energy $f = - \ln Z$, and we are to compute the disorder average
$\langle f \rangle$.  To tackle that task, it is popular among
theoretical physicists to follow a recipe called the replica trick.
Instead of attempting to calculate $\langle \ln Z \rangle$ directly,
one computes the disorder average $f_n \equiv \langle Z^n \rangle$ for
all positive integers $n = 1, 2, ... \,\infty$, which is be done by
introducing $n$ copies, or replicas, of the system.  Then, recalling
the identity $\ln Z = \lim_{n\to 0} (Z^n - 1) / n$, and assuming some
extension of the discrete set $(f_n)_{n \in {\mathbb N}}$ to an
analytic function $f(u)$ on ${\mathbb C}$, one hopes that $f^\prime
(0)$ equals $-\langle \ln Z \rangle$.  An obvious problem with this
trick is the non-uniqueness of the analytic continuation.  Indeed, if
$f(u)$ satisfies $f(u) \big|_{u=n} = f_n$, then so does $f(u) + c /
\Gamma(-u)$, where $\Gamma(u)$ is the Euler gamma function and $c$ is 
an arbitrary constant.  Notice that the increment $c / \Gamma(-u)$
vanishes for all positive integers $u = n$ but has finite derivative
at $u = 0$.  Thus we do not know whether the answer for $- \langle \ln
Z \rangle$ is $f^\prime(0)$ or $f^\prime(0) - c$.  To determine the
unknown constant, which may be a function of temperature, disorder
strength {\it etc.}, one must have some control on $\langle Z^u
\rangle$, for example a bound on its asymptotic behavior as $u \to
i\infty$.  Sadly, such control is not always available.

Some time ago, Verbaarschot and the author (VZ) addressed \cite{vz} the 
above concern in the context of disordered electron systems, {\it i.e.}, 
the problem of computing averages of retarded and advanced Green's
functions $G^\pm(E) = (E\pm i\epsilon - H)^{-1}$ of a disordered
single-particle Hamiltonian $H$.  For such systems, one implements the
replica trick by starting from a generating function
	$$
	Z_{m,n}(\omega,J_1,J_2) = \Big\langle 
	{\rm Det}^m(E+\omega+i\epsilon-H+J_1) \, 
	{\rm Det}^n(E-i\epsilon-H+J_2) \Big\rangle 
	$$ 
depending on matrix sources $J_1, J_2$.  Angular brackets denote
averaging over disorder. For positive integers $m$ and $n$, this
generating function can be expressed as a Gaussian integral over
anticommuting complex fields (``fermionic replicas''), and for
negative integers over commuting fields (``bosonic replicas'').  VZ
focused on the spectral two-point function,
	$$
	S_2(\omega) = \Big\langle
	{\rm Tr} (E+\omega+i\epsilon-H)^{-1} \ 
	{\rm Tr} (E-i\epsilon-H)^{-1} \Big\rangle
	\quad (\epsilon > 0),
	$$
for the simple case of Wigner-Dyson statistics for $\beta = 2$, {\it
i.e.}, the Gaussian ensemble of complex Hermitian matrices with
unitary symmetry (GUE).  By following the standard procedure
\cite{wegner1,elk,efetov} of mapping the problem on a nonlinear 
$\sigma$ model, and then doing the natural analytic continuation $m=n
\to 0$, VZ found that fermionic and bosonic replicas give different
answers, and both answers differ from the exact result.  What is
reproduced correctly by the replica trick is the asymptotics of
$S_2(\omega)$ as $\omega \to \infty$.  This finding led to a general
consensus that the replica trick for disordered electron systems is
limited to those regions of parameter space where the nonlinear 
$\sigma$ model can be evaluated perturbatively.

This consensus is now being challenged by recent proposals
of Kamenev and Mezard \cite{km1,km2}, and of Yurkevich and Lerner
\cite{yl}.  A partial list of their claims is as follows:  i) 
According to Kamenev and Mezard \cite{km1}, past work using the
replica trick for disordered electron systems suffers from the tacit
assumption of the absence of ``spontaneous breaking of replica
symmetry''.  ii) When replica-symmetry broken saddle points are
included, the ``standard'' nonlinear $\sigma$ model derived from
fermionic replicas gives the correct oscillatory asymptotics of the
level correlation functions.  iii) Perturbation theory in
$\omega^{-1}$ yields a ``systematic`` expansion for the amplitudes of
the oscillatory terms.  iv) The fermionic replica trick with
replica-symmetry breaking is ``nonperturbative'' and reproduces the
results of the supersymmetric method.  v) Yurkevich and Lerner
speculate that if limits are taken in the proper order (first $n \to
0$, then $\omega \to 0$), the fermionic replica trick may reproduce
also the {\it singular} part of the two-point function.

These claims and speculations add up to a substantial revision of the
accepted picture.  They provoke some sort of response, and it is the
purpose of the present paper to offer clarification.  Section
\ref{sec:one-point} calculates in some detail the one-point function,
Section \ref{sec:two-point} reviews the two-point function, Section
\ref{sec:DH} explains why replicas can be manipulated to yield the
exact answer for the spectral correlations of the ${\rm GUE}$, and
Section \ref{sec:summary} summarizes the conclusions. 

\newpage
\section{Fermionic replicas: one-point function}
\label{sec:one-point}

Let us start from the beginning and consider first the spectral
one-point function $S_1(z) = \langle {\rm Tr} \, (z-H)^{-1}\rangle$.
The fundamental difficulty plaguing fermionic replicas for disordered
electron systems is already visible in this case.

On elementary grounds, the function $S_1(z)$ has a branch cut along
the real axis (or the line of spectral support of the Hamiltonian $H$), 
and its discontinuity across the cut yields the average density of 
states, $\rho(E)$:
	$$
	\rho(E) = {1 \over 2\pi i} \lim_{\epsilon\to 0+}
	\Big( S_1(E-i\epsilon) - S_1(E+i\epsilon) \Big) 
	= {1 \over \pi} \lim_{\epsilon\to 0+} {\rm Im}
	\, S_1(E-i\epsilon) \;.
	$$
The replica trick, instead of attempting to calculate $S_1$ directly, 
looks at the discrete family of functions
	$$
	f_n(z) = \Big\langle {\rm Tr} \, (z-H)^{-1} 
	{\rm Det}^n(z-H) \Big\rangle \qquad (z \in {\mathbb C})
	$$
for $n \in {\mathbb Z}$.  The member $f_0(z)$ of this family is the
desired quantity $S_1(z)$.  In the {\it fermionic} version of the
replica trick, one computes $f_n(z)$ for all {\it positive} integers
$n = 1, 2, \ldots \infty$, and hopes to infer $S_1(z)$ by setting $n =
0$ at the end of the calculation.  On brief reflection, however, this
hope must appear ill-founded.  Indeed, it is easy to see that, if all
moments $\{ \langle {\rm Tr} \, H^p \rangle \}_{p = 1, \ldots,
n-1}$ exist, the function $f_n(z)$ is {\it analytic} in $z$ for all
positive integers $n \in {\mathbb N}$, which means in particular that
$f_n(z)$ is continuous across the real axis, and
	$$
	\rho_n(E) \equiv {1 \over 2\pi i} \lim_{\epsilon\to 0+}
	\Big( f_n(E-i\epsilon) - f_n(E+i\epsilon) \Big) 
	= {1 \over \pi} \lim_{\epsilon\to 0+} {\rm Im}
	\, f_n(E-i\epsilon) = 0
	$$
is zero, for all $E \in {\mathbb R}$ and $n \in {\mathbb N}$.  This is
beginning to look bizarre.  Given that all imaginary parts $\rho_n(E)$
vanish identically for $n = 1, 2, \ldots \infty$, how can we hope to
predict, based on these data only, the nonzero value of the average
density of states $\rho(E) =
\rho_n(E)\big|_{n=0}$?!

Let us postpone the answer for a moment and jump to the computation of
$f_n(z)$ $(n \ge 1)$ using fermionic replicas.  For simplicity we
focus on the case of $N \times N$ random matrices with $\beta = 2$
(unitary symmetry) and take the probability measure for the complex
Hermitian matrix $H$ to be Gaussian with weight function $P(H) = {\rm
const} \times \exp(- N {\rm Tr}H^2 / 2w^2)$.  By a standard sequence
of transformations \cite{elk,vz}, the function $f_n(z)$ can then be 
cast in the form of an integral over ${\rm u}(n)$, the Lie algebra of 
the unitary group ${\rm U}(n)$:
	$$
	f_n(z) = {N \over n} \int_{{\rm u}(n)} {\rm Tr} \, (z-Q)^{-1} \,
	{\rm Det}^N (z-Q) \, {\rm e}^{N {\rm Tr} \, Q^2 / 2 w^2} dQ \;,
	$$
where $dQ$ is the flat measure on ${\rm u}(n)$ normalized by $\int_{
{\rm u}(n)} {\rm e}^{N {\rm Tr} \, Q^2 / 2w^2} dQ = 1$.  The elements
of ${\rm u}(n)$ are skew-Hermitian matrices, whence ${\rm Tr} \, Q^2 =
- {\rm Tr} \, Q^\dagger Q \le 0$ and the integral exists.  Note that
the integrand is invariant under the adjoint action $Q \mapsto U Q
U^{-1}$ of ${\rm U}(n)$ on its Lie algebra.

For large $N$, the integral over $Q$ is saturated by the solutions of
the saddle point equation
	$$
	{\delta \over \delta Q} \left( {\rm Tr} \, Q^2 / 2 w^2 + 
	{\rm Tr} \ln (z-Q) \right) = 0 = Q / w^2 + (Q-z)^{-1} \;.
	$$
This equation is of real type.  Its solutions therefore come in 
Hermitian conjugate pairs: if $Q$ is a solution for $z=E\in{\mathbb 
R}$, then so is $Q^\dagger$.  In particular, for $|E| < 2w$ there 
exists a solution
	$$
	Q =  w {\rm e}^{i\theta} \times 1_n \;, \qquad 
	{\rm e}^{i\theta} = E/2w + i \sqrt{1 - (E/2w)^2} \;,
	$$
which is proportional to the unit matrix, and is paired with $Q^
\dagger = w {\rm e}^{-i\theta} \times 1_n$.  These two solutions are
unique in that they are stable under the adjoint action of ${\rm
U}(n)$.  For $n \ge 2$, there exist also solutions with less symmetry,
which organize into smooth saddle-point manifolds.  They can be
constructed by starting from a diagonal solution with $n-p$ (resp.
$p$) matrix elements equal to $w {\rm e}^{i\theta}$ (resp.  $w {\rm
e}^{-i\theta}$), and then applying the adjoint action of the symmetry
group ${\rm U}(n)$.

The point to be made here is this: proper application of the
saddle-point method to the integral representation for $f_n(z)$
requires us {\it to sum over pairs of conjugate saddles}.  It is clear
that the contribution from a saddle point (or saddle-point manifold)
$Q$ is precisely the complex conjugate of the contribution from its
Hermitian conjugate $Q^\dagger$, if $z = E \pm i 0$.  Therefore, we
can be sure that the result for $f_n(z)$ on the real energy axis is
{\it real}:
	$$
	f_n(E) = \varphi_n(E) + \bar\varphi_n(E) \in {\mathbb R} \;,
	$$
and $\rho_n(E) = \pi^{-1} \lim_{\epsilon\to 0} f_n(E-i\epsilon)$
is zero for all $n = 1, 2, ... \infty$.  This, of course, was to
be expected in view of what we saw at the beginning of the section. 

We now return to the question posed before.  Since $\rho_n(E)$
vanishes identically for all $n \in {\mathbb N}$, how can we manage to
produce a reasonable result for $\rho(E) = \rho_n(E)\big|_{n=0}$?
Clearly, the answer is that some additional input must be injected.
What is missing from the formalism with fermionic replicas is the
entire information about causality, {\it i.e.}, the distinction
between retarded and advanced Green's functions ($z = E + i \epsilon$,
versus $z = E - i \epsilon$).  This information is present in $f_n(z)$
only for $n \le 0$.  In the bosonic replica trick, where one evaluates
$f_n(z)$ for $n = -1, -2, \ldots -\infty$, causality is reflected in
the $Q$-integral representation as a singularity of the integrand,
which restricts the number of saddles that are accessible by
continuous deformation of the integration contour \cite{sw}.  It turns
out that for all $n < 0$, there exists just one saddle point which is
accessible, and all others are inaccessible.  On the other hand, in
the rigorous formulation of $f_0(z)$ by the supersymmetric method
\cite{brezin}, there are four saddle points, two accessible and two
inaccessible ones.  Both the supersymmetric formula and bosonic 
replicas will be discussed in more detail below.

\subsection{The recipe}
\label{sec:recipe}

As we have seen, the reason for the vanishing of $\rho_n(E)$ $(n \in
{\mathbb N})$ is the {\it pairwise} appearance of saddles as conjugate
pairs $(Q , Q^\dagger )$.  It is therefore clear that, in order to
assist the fermionic replica trick and extrapolate $\rho_n(E)$ to a
nonvanishing answer for $n = 0$, we must come up with a recipe for
breaking the symmetry between the members of conjugate pairs.  Let us
denote by $Q_p$ the diagonal matrix with the first $n-p$ entries equal
to $w {\rm e}^{i\theta}$ and the last $p$ entries equal to $w {\rm
e}^{-i\theta}$.  If $z = E \in {\mathbb R}$, the matrix $Q_p^\dagger$
lies on the saddle-point manifold $U Q_{n-p} U^{-1}$ $(U \in {\rm
U}(n))$, so the pairing of saddles is $p \leftrightarrow n-p$.

For definiteness, let us fix $z = E - i0$, in which case ${\rm Im} \,
S_1(z) \ge 0$.  According to the saddle-point equation $(z-Q)^{-1} = 
Q/w^2$, the expression
	$$
	{\rm Im} \, (E - w {\rm e}^{i\theta})^{-1} = {
	\rm Im} \, {\rm e}^{i\theta} / w = w^{-1} \sqrt{1-(E/2w)^2}
	$$
is positive for $|E| < 2w$, which is known \cite{mehta} to be the
range of the energy spectrum in the limit $N \to \infty$.  Since the
integral representation for $f_n(z)$ contains the factor ${\rm Tr} \,
(z-Q)^{-1}$, the ``good'' saddle point, contributing to $f_n(z)$
with an imaginary part of the desired positive sign, is $Q_0 = w {\rm
e}^{i\theta} \times 1_n$.  By the same token, $Q_n = w {\rm e}^ {-
i\theta} \times 1_n$ is a ``bad'' saddle point.  Alternatively, we
could say that $Q_0$ is ``causal'', whereas $Q_n$ is ``acausal''.
By heritage, or continuity, the natural recipe now is to declare the
saddle-point manifolds $U Q_p U^{-1}$ with $p = 0, 1, 2, \ldots$ to 
be ``causal'', and their conjugates $p = n, n-1, n-2, \ldots$ to be 
``acausal''. (Clearly, this only makes sense if there is some mechanism 
that terminates the sequences before they overlap.)  Anticipating the
emergence of causality for $n = 0$, we then retain only the
contributions from the ``causal'' saddles, while throwing away the
others.  The resulting fake expressions for $f_n(E-i\epsilon)$ have 
nonzero imaginary parts, and extrapolation to $n = 0$ can now give a 
finite density of states.  The result so obtained for $|E| < 2w$ 
reads \cite{km1}
	\begin{eqnarray}	
	\rho &=& (N/\pi w) \sin\theta + {\cal O}(N^0) + ... \nonumber\\
	&&- (4\pi w \sin^2 \theta)^{-1} \cos \Big( N (2\theta - 
	\sin 2\theta) \Big) + {\cal O}(N^{-1}) + ... \;.
	\label{asymptotics}
	\end{eqnarray}
The leading term is Wigner's famous semicircle law, $\sin\theta = 
\sqrt{1-(E/2w)^2}$, and stems from the invariant saddle point $p = 0$. 
The subleading (oscillatory) term originates from the saddle-point
manifold $p = 1$, which is generated by applying the adjoint action of
${\rm U}(n)$ to $Q_1$, the diagonal matrix with $n-1$ entries $w {\rm
e}^{i\theta}$ and one entry $w {\rm e}^{-i\theta}$.  The stability
group of $Q_1$ in ${\rm U}(n)$ is ${\rm U}(n-1) \times {\rm U}(1)$.
Hence the saddle-point manifold $p = 1$ is isomorphic to the quotient
${\rm U}(n) / {\rm U}(n-1) \times {\rm U}(1)$.  More generally, the 
saddle-point manifold with index $p$ is isomorphic to the Grassmannian
$G_{n,p} = {\rm U}(n) / {\rm U}(n-p) \times {\rm U}(p)$, which is a
symmetric space of type $A{\rm III}$ in Cartan's classification
\cite{helgason}.  Carrying out the saddle-point approximation requires 
integrating over the directions transverse to the saddle-point
manifold approximately, and over the saddle-point manifold itself
exactly.  The latter integral produces a volume factor ${\rm vol}
(G_{n,p})$.  For $p = 1$, this volume goes essentially as $\Gamma(n)
^{-1}$ (see for example the appendix of \cite{km2}), which combines
with the prefactor $n^{-1}$ of the integral representation for $f_n(z)$ 
to produce a finite limit $n^{-1} \Gamma(n)^{-1} \to 1$ for $n = 0$.  
More generally, ``causal'' analytic continuation \cite{km1} of the volume 
of $G_{n,p}$ yields a power $n^p$ near $n = 0$.  The contributions from 
the saddle-point manifolds $p \ge 2$ therefore extrapolate to zero 
at $n = 0$.  

Remarkably, the above $N^{-1}$ expansion of the average density of
states agrees \cite{km1} with the asymptotic behavior that can be
inferred from an exact representation in terms of orthogonal
polynomials.  In view of the rather ad hoc nature of the derivation,
mathematicians will throw up their hands in horror or despair, while
physicists are much intrigued.  In any case, putting aside questions
of mathematical rigor, it is definitely desirable to gain a better
understanding of {\it why} the procedure works.

\subsection{Supersymmetry}

One avenue towards some understanding is to make a comparison with the
rigorous representation \cite{brezin,mrz_suprev} of $S_1(z)$ as an
integral over a $2 \times 2$ supermatrix $Q = \pmatrix{x &\xi\cr \eta
&iy \cr}$:
	$$
	S_1(z) = {N \over 2} \int DQ \, {\rm Tr} \, (z-Q)^{-1} \,
	{\rm SDet}^{-N} (z-Q) \, {\rm e}^{-N {\rm STr} \, Q^2 / 2 w^2} \;.
	$$
Here $DQ$ stands for the flat Berezin form $DQ = (2\pi)^{-1} {\rm d}x
{\rm d}y \, {\partial^2 / \partial\xi\partial\eta}$, and the integral is
over $x \in {\mathbb R}$ and $iy \in i{\mathbb R}$.  The symbols ${\rm
STr}$ and ${\rm SDet}$ denote the supertrace and the superdeterminant.

The above Berezin integral for $S_1(z)$ has a total of four saddle
points, obtained by going through the various sign choices in $Q = w
\, {\rm diag}({\rm e}^{\pm i\theta}, {\rm e}^{\pm i\theta})$.  Two of
these are ``bad'', as they cannot be reached by continuous deformation
of the integration contour for $x$ without crossing the $N$-th order
pole of ${\rm SDet}^{-N} (z-Q)$ at $z = x$.  As before, let us fix $z 
= E-i0$.  Then the accessible saddle points are $Q_0 = w \, {\rm diag}
( {\rm e}^{i\theta} , {\rm e}^{i\theta} )$ and $Q_1 = w \, {\rm diag}
( {\rm e}^{i\theta} , {\rm e}^{-i\theta} )$.  The first one gives rise
\cite{brezin} to Wigner's semicircle, and on comparing with the
fermionic replica trick, one begins to suspect that perturbation
theory around the second one might yield the oscillatory contribution
to the asymptotic result (\ref{asymptotics}).  I have checked that
this is precisely the case.

In hindsight, the coincidence is not all that surprising.  At the
non-invariant saddle point $Q_1 = w \, {\rm diag}({\rm e}^{i\theta} ,
{\rm e}^{-i\theta})$, both $\xi$ and $\eta$ are Goldstone fermions of
the supersymmetric integrand ${\rm SDet}^{-N}(z-Q) \exp - N {\rm STr}
\, Q^2 / 2w^2$.  On the other hand, the $p = 1$ saddle-point manifold
${\rm U}(n) / {\rm U}(n-1) \times {\rm U}(1)$ of the fermionic replica
trick is isomorphic to the complex projective space ${\mathbb C}{\rm
P}_{n-1}$, which is a manifold of real dimension $2(n-1)$. At $n = 0$,
this has real dimension $-2$, and has to be understood as a space
generated by two anticommuting degrees of freedom, say $\xi$ and
$\eta$.  Hence, integrating over ${\mathbb C}{\rm P}_{n-1}
\simeq {\rm U}(n) / {\rm U}(n-1) \times {\rm U}(1)$ and then 
setting $n = 0$ is just a complicated way of taking two derivatives
$\partial^2 / \partial\xi \partial\eta$.  The latter is what the
Berezin integral formula does.  The agreement between the asymptotic
expansions to leading order motivates us to propose the following
conjecture: the fermionic replica trick, augmented by the plausible
recipe of retaining only the saddles $p = 0$ and $p = 1$, is {\it
perturbatively equivalent} to the supersymmetric integral
representation for $S_1(z)$.  In other words, pushing the asymptotic
expansion around the two saddles $Q_0$ and $Q_1$ to higher order, we
expect to get agreement in every order of $N^{-1}$.

Can this conjecture be put on a more solid footing?  With the goal of
justifying better the asymptotic expansion obtained from the fermionic
replica trick, we are now going to take a more close-up look at the
procedure of analytic continuation in $n$.

\subsection{Analytic continuation?}

To keep the calculations as simple as possible, we start from the
generating function
	$$
	F_n(z) = \Big\langle {\rm Det}^{n}(z-H) \Big\rangle \;,
	$$
aiming to eventually apply $n^{-1} d/dz$ and analytically continue to
$n = 0$, so as to extract $S_1(z)$.  As we have seen, the values of
$F_n(z)$ at the positive integers $n$ do not contain sufficient
information to permit analytic continuation.  Moreover, it is not
known how to compute $F_n(z)$ for $n \notin {\mathbb Z}$, at not least
not directly.  We {\it do} have access, however, to $F_n(z)$ at the
negative integers $n$, by means of the {\it bosonic} replica trick.
Let us see what additional information we can collect from there.

By some elementary transformations, we arrive at the identity
	$$
	F_{-n}(z) = 
	\int_{i{\rm u}(n)} {\rm Det}^{-N} (z-Q) \, 
	{\rm e}^{-N {\rm Tr} \, Q^2 / 2 w^2} dQ 
	\qquad (n \in {\mathbb N})\;.
	$$
We are now integrating over the space of {\it Hermitian} matrices,
viewed as $i{\rm u}(n)$.  The saddle-point equation governing the
large-$N$ limit still reads $(z-Q)^{-1} = Q/w^2$ and, as before, the
invariant $p = 0$ saddle point yields \cite{es,vz1} Wigner's
semicircle.  Other saddle-point manifolds with index $p = 1, 2, ...,
n$ exist, too.  The difference from before is that these are now {\it
disconnected} from the integration domain $i{\rm u}(n)$ by the
singularities of the function ${\rm Det}^{-N}(z-Q)$.  In view of what
happend for fermionic replicas, one might think that one should
ignore disconnectedness and incorporate the other saddles.  Surely,
moving across the $N$-th order pole of ${\rm Det}^{-N}(z-Q)$ just
produces some residue, and one could argue that this is schematically
given by $\delta^{N-1} / \delta Q^{N-1} \exp (- N {\rm Tr} \, Q^2 /
2w^2) \big|_{Q = z}$, which is exponentially small and hence
negligible, for large $N$.  However, we are on the wrong track.  A
little thought shows that the saddles $p = 0, 1, ...,n$ are aligned 
{\it transversely} to the contour of integration for $Q$. This makes 
it impossible to deform the contour so as to pass through more than
one member of the sequence of saddles without backtracking. (In
contrast, for fermionic replicas the alignment is {\it longitudinal},
and a constant shift $Q \to Q + E/2$ puts all the saddles right on the
contour.)  Moreover, if one proceeds anyway and computes the
contribution from $p = 1$, one obtains an expression containing the
exponential factor 
	$$
	\exp \left( nN(i\theta - {\rm e}^{i\theta}/2) - iN
	(2\theta - \sin 2\theta) \right) \;.
	$$
(Again, this is for $z = E - i0$.)  Near $n = 0$ this can be seen to
{\it increase without bound} as $z$ moves away from the real axis toward 
$z = -i\infty$.  Such behavior is incompatible with analyticity of the
Green's function at infinity.  For all these reasons, it is safe to 
say that $F_n(z)$ at the negative integers $n$ is exhausted by a {\it 
single} saddle point, $p = 0$.  Evaluation of the corresponding integral 
in Gaussian approximation gives
	\begin{equation}
	F_n(2w\cos\theta - i0) = w^{nN} \, 
	{ \exp \left( nN(\textstyle{1\over 2} {\rm e}^{2i\theta}
	- i\theta) \right) \over \sqrt{1 - {\rm e}^{2i\theta}}^{\,n^2} } 
	+ {\cal O}(N^{-1}) \quad (-n \in {\mathbb N}) \;.	
	\label{bosonic}
	\end{equation}
This result will place a strong constraint on the analytic continuation
of $F_n$.  

Next we return to fermionic replicas, and carry out a more complete
saddle-point evaluation of 
	$$
	F_n(z) = \int_{{\rm u}(n)} {\rm Det}^{N} (z-Q) \, 
	{\rm e}^{N {\rm Tr} \, Q^2 / 2 w^2} dQ 
	$$
for $n \in {\mathbb N}$.  As before, let $Q_p$ denote a diagonal
matrix with $n-p$ entries $w {\rm e}^{i\theta}$ and $p$ entries $w
{\rm e}^{-i\theta}$.  By acting with the symmetry group ${\rm U}(n)$
on $Q_p$ we get the saddle-point manifold $G_{n,p} = {\rm U}(n) / {\rm
U}(n-p) \times {\rm U}(p)$.  To perform the saddle-point approximation, 
we need to factor the integration domain near $G_{n,p}$ in a suitable
manner.  For this purpose, let ${\rm u}(n)$ be decomposed into two
linear subspaces: ${\rm u}(n) = V + W$, where the elements of $V =
{\rm u}(n-p) + {\rm u}(p)$ commute, while those of $W$ anticommute, 
with $Q_p$.  We then introduce adapted coordinates by
	$$
	Q = U ( Q_p + X ) U^{-1} \;,
	$$
where $U Q_p U^{-1}$ parametrizes $G_{n,p}$, and $X \in V$ spans the
directions transverse to $G_{n,p}$.  By an elementary calculation,
the volume element $dQ$ under this factorization transforms into
	$$
	{\rm Det} \, {\rm ad}(Q_p + X)\big|_W \, dX \, dU \;.
	$$
Here, $dU$ is a ${\rm U}(n)$-invariant measure for $G_{n,p}$, $dX$ is
a flat measure on $V$, and ${\rm ad}(Y)\big|_W$ denotes the commutator
(or adjoint) action of $Y \in V+iV$ on the elements of $W+iW$.  By
expanding the integrand ${\rm Det}^N(z-Q) \exp N {\rm Tr}\,Q^2/2w^2$ 
w.r.t. $X$ and doing the integral over $X$ in Gaussian approximation, 
we obtain
	\begin{eqnarray}
	F_n &\simeq& \sum_{p=0}^n {\rm Det}^{-N}(Q_p / w^2) \, 
	\exp \left( N {\rm Tr} \, Q_p^2 / 2w^2 \right) \, 
	{\rm Det} \, {\rm ad}(Q_p) \big|_W \nonumber \\
	&&\times \int_{G_{n,p}} dU \int_V {\rm e}^{(N/2w^2) \, 
	{\rm Tr}\, (X^2 - w^{-2} X Q_p X Q_p)} dX \nonumber \\
	&=& w^{nN} \sum_{p=0}^n {\rm vol}(G_{n,p}) (N / 2\pi)^{p(n-p)}
	\, { ({\rm e}^{i\theta} - {\rm e}^{-i\theta})
	^{2p(n-p)} \over \sqrt{1-{\rm e}^{2i\theta}}^{\,(n-p)^2}
	\sqrt{1-{\rm e}^{-2i\theta}}^{\,p^2} } \nonumber \\
	&&\times \exp \left( nN(\textstyle{1\over 2} {\rm e}^{2i\theta}
	- i\theta) + ipN (2\theta - \sin 2\theta) \right) 
	\qquad (n \in {\mathbb N}) \;.
	\label{fermionic}
	\end{eqnarray}
An analogous expression for the derivative $n^{-1} dF_n / dz$ was
obtained in \cite{km1}.  We now face the task of finding an analytic
continuation of $F_n$ to $n \in {\mathbb C}$ which respects the
results (\ref{bosonic}) and (\ref{fermionic}) for $n \in {\mathbb Z}$.
Kamenev and Mezard \cite{km1} have suggested that the solution to
this problem is to analytically continue ${\rm vol}(G_{n,p})$ as
	$$
	{\rm vol} \left( G_{n,p} \right) = (2\pi)^{p(n-p)}
	\prod_{j = 1}^p {\Gamma(j) \over \Gamma(n-j+1)} \;,
	$$
and extend the upper limit on the sum over $p$ in (\ref{fermionic}) to
infinity.  The latter does not change the value of $F_n$ for $n \in
{\mathbb N}$, as $\Gamma(n-p+1)^{-1}$ vanishes for $p \ge n+1$.  For
the same reason, the terms $p \ge 1$ vanish at the negative integers
$n$, in agreement with the constraint posed by (\ref{bosonic}).  So
far so good.  Unfortunately, when $n$ is not an integer, the infinite
sum over $p$ is ill-behaved.  The culprit is the rapid growth with $p$
of the (analytically continued) volume factors:

	$$
	(2\pi)^{2p-n-1} {{\rm vol}(G_{n,p}) \over {\rm vol}(G_{n,p-1})}
	= {\Gamma(p) \over \Gamma(n-p+1)} = 
	\Gamma(p) \Gamma(p-n) \, \pi^{-1} \sin\pi (p-n) \;, 
	$$
which makes the sum over $p$ diverge.  There appears to exist no
obvious remedy for this difficulty, and we have to concede that
extension of the sum over $p$ to infinity does {\it not} define
$F_n(z)$ away from the integers.  Thus we have so far failed in our
attempt to perform a bona fide analytic continuation, and the recipe
of Section \ref{sec:recipe} still awaits justification.

Nevertheless, there seems to be some truth to the proposal by Kamenev
and Mezard.  If we boldly ignore the divergence issue, and formally
apply $(\pi n)^{-1} d/dz$ at $n = 0$, then all the terms for $p \ge 2$
disappear, and the answer from $p = 0$ and $p = 1$ can be shown to
agree with the large-$N$ asymptotics extracted from the exact formula
for $S_1(z)$ as a Berezin integral.  Upon further thought, the failure
of the proposed analytic continuation can be interpreted as follows.
Recall that the failure is caused by the late terms in the sum over
$p$, which in turn stem from our insisting that the analytic
continuation reproduce the result (\ref{fermionic}) for {\it all} $n
\in {\mathbb N}$.  But at fixed $N < \infty$, the saddle-point
approximation leading to that expression {\it breaks down} when $n$
becomes large.  Indeed, for $n \gg N$ the $Q$-integral formula for
$F_n(z)$ is inappropriate, and it is better to return to the original
representation
	$$
	F_n(z) = \int {\rm Det}^n(z-H) \, {\rm e}^{-N {\rm Tr}
	H^2 / 2w^2} dH \;.
	$$
By evaluating this integral in saddle-point approximation, one gets a
an answer formally similar to (\ref{fermionic}).  A major difference
is that the sum over saddle-point manifolds $p$ now terminates at
$N$, not $n$.  Thus, making $n$ bigger no longer extends the range of
$p$.  The lesson to be learnt from this is that the approximate result
(\ref{fermionic}) must not be trusted for large values of $n$.  It is 
likely that the divergence of the sum over $p$ is fake and is somehow
cut off when a more correct approximation for large $n$ is used.  When 
$N$ is big enough, this cutoff will have a negligible effect on the 
extrapolation to $n = 0$.  On the other hand, for small values of $N$ 
the cutoff will probably lead to some unknown correction.

We could have anticipated the existence of such corrections already
from our simplified presentation in Section \ref{sec:recipe}.  Surely,
the recipe of throwing away half of the saddles makes sense only in
the perturbative domain of isolated saddles.  When the saddles are not
well separated (as is the case for small $N$), their asymptotic series
in some sense ``interfere'', which is likely to give rise to
corrections of exponentially small type $\sim {\rm e}^{- {\rm const}
\times N}$.  For the one-point function, the existence of such 
corrections is of minor concern, as we are usually satisfied with
knowing the large-$N$ asymptotics.  As we shall see, however, the
situation is less favorable for the two-point function.  Note also
that the question of what happens beyond the saddle-point
approximation or, more precisely, whether a non-analytic function can
be reconstructed from its asymptotic series (by Borel resummation, for
example), is a very difficult issue.

Let me finish the section with a word on vocabulary.  In my opinion it
is inappropriate to call the appearance of the $p = 1$ saddle-point
manifold an instance of ``spontaneous breaking of replica symmetry''.
Spontaneous symmetry breaking, a phenomenon which may occur in
physical systems with infinitely many degrees of freedom, means that a
symmetry of the Hamiltonian is not manifest in the (ground) state of
the system.  The symmetry group of the present problem is ${\rm
U}(n)$.  The saddle-point manifold ${\mathbb C}{\rm P}_{n-1} \simeq
{\rm U}(n) / {\rm U}(n-1) \times {\rm U}(1)$ is a beautiful symmetric
(and replica-symmetric) space, whose geometry is invariant under the
action of ${\rm U}(n)$.  It is true, of course, that an {\it
individual} point on the manifold is not invariant under ${\rm U}(n)$.
However, our job is to integrate, and therefore the issue is not
invariance of individual points but invariance of the {\it domain of
integration}.  The integration manifold ${\mathbb C}{\rm P}_{n-1}
\simeq {\rm U}(n) / {\rm U}(n-1)\times {\rm U}(1)$ definitely does
{\it not} break ${\rm U}(n)$ replica symmetry, just as the familiar
two-sphere in Euclidean 3-space, ${\rm S}^2 = {\rm SO}(3) / {\rm
SO}(2)$, does not break ${\rm SO}(3)$ rotational symmetry.

\section{Fermionic replicas: two-point function}
\label{sec:two-point}

Let $G$ be one of the compact matrix groups ${\rm Sp}(2m+2n)$, ${\rm
U}(m+n)$, or ${\rm O}(2m+2n)$.  Acting with $G$ on a fixed matrix
$\Sigma_3$ by conjugation, we get an orbit of elements we denote by
$Q = U \Sigma_3 U^{-1}$ $(U \in G)$.  For $G = {\rm Sp}, {\rm O}$ we
take $\Sigma_3 = {\rm diag}(1_{2m} \, , -1_{2n})$; and for $G = {\rm
U}$, $\Sigma_3 = {\rm diag}(1_m \, , -1_n)$.  To keep the dimensions
explicit, we write $X_{m,n}$ for the orbit of $G$ on $\Sigma_3$.  The
stability group of $\Sigma_3$ is a subgroup, $H$, of $G$.  Letting $G$
act on $\Sigma_3$ by conjugation amounts to taking the quotient of $G$
by $H$, so $X_{m,n}$ is isomorphic to the coset space $G/H$ (which is
moreover a symmetric space).

Now let $dQ$ be a $G$-invariant measure on $X_{m,n}$, and consider
the generating function
	$$
	Z_{m,n}(\omega) = \int_{X_{m,n}} \exp \left(
	{i\omega\over 2\alpha} {\rm Tr} \, \Sigma_3 Q \right) dQ \;.
	$$
This is the zero-dimensional limit of the nonlinear $\sigma$ model of
Wegner \cite{wegner1} and Efetov \cite{efetov}, truncated to the
compact sector arising from replicated fermionic fields \cite{elk}.
The generating function for $G = {\rm Sp}, {\rm U}, {\rm O}$ is
intended to describe the universal low-frequency limit of a disordered 
electron system with symmetry index $\beta = 1, 2, 4$ ({\it i.e.},
orthogonal, unitary, or symplectic symmetry), in this order.  The
dimensionless parameter $\omega$ is a frequency measured in units of
the mean level spacing divided by $\pi$, and $\alpha = 2,1,1$ for
$\beta = 1,2,4$.  For future use, note that the integrand is invariant
under the action of the little group $H$.

The transformation $Q \mapsto -Q$ is a natural isomorphism from
$X_{m,n}$ to $X_{n,m}$.  Using it in the expression for $Z_{m,n}
(\omega)$ we obtain the symmetry relation
	$$
	Z_{m,n}(\omega) = Z_{n,m}(-\omega).
	$$
This symmetry does not come as a surprise but reflects the fact
that $Z_{m,n}(\omega)$ is a low-frequency approximation to
	$$
	\Big\langle {\rm Det}^m(E-i\epsilon+\omega/2-H) \, 
	{\rm Det}^n(E+i\epsilon-\omega/2-H) \Big\rangle \;.
	$$
Because this is an {\it analytic} function of $\omega/2-i\epsilon$ for
$m,n\in{\mathbb N}$, nothing is lost by setting $\epsilon = 0$.  We
then see at a glance that the generating function is invariant under the
combined operation of exchanging $m \leftrightarrow n$ and sending
$\omega \to -\omega$.

Recall the definition of the two-point function 
	$$
	S_2(u,v) = \Big\langle {\rm Tr} \, (u-H)^{-1}
	\, {\rm Tr} \, (v-H)^{-1} \Big\rangle \;.
	$$
We are interested in the case where the arguments of $S_2$ lie on
opposite sides of the real energy axis: $u = E_1-i0$, $v = E_2+i0$,
and their difference is a fixed multiple $\omega/\pi$ of the mean level
spacing.  The resulting function is still denoted by $S_2(\omega)$.

In the fermionic replica trick for the two-point function, one hopes
to extract $S_2(\omega)$ from the knowledge of $Z_{m,n}(\omega)$ for
all positive integers $m$ and $n$, by applying two derivatives $-
(mn)^{-1} \partial^2 / \partial\omega^2$ and setting $m = n = 0$.  The
bad news is that this hope is frustrated by the symmetry relation
$Z_{m,n}(\omega) = Z_{n,m}(-\omega)$.  If the replica limit exists in
a naive sense, then it cannot matter whether we first set $m = 0$ and
then $n = 0$, or the other way around.  Another option is to set $m -
n$ = 0 first and $m + n = 0$ afterwards.  If we proceed in the latter
fashion, the symmetry of the generating function reduces to
$Z_{n,n}(\omega) = Z_{n,n}(-\omega)$.  Thus, $Z_{n,n}(\omega)$ is an
even and hence real function of $\omega$.  By naive extrapolation to
$n = 0$, we would conclude that $S_2(\omega)$ is an even and real
function of its argument.  But this clearly is nonsense, for $S_2$ is
neither even nor real.  For example, if ${\rm Im}\,\omega<0$ the
universal answer for the case $\beta = 2$ is $S_2(\omega) = 1 -
2i\omega^{ -2} {\rm e}^{-i\omega} \sin\omega$.

The reason for the nonsensical answer produced by our use of fermionic
replicas is the same as for the one-point function: the analyticity in
$\omega$ of the function $Z_{m,n}(\omega)$ means that the fermionic
replica trick suffers from the deficiency of being entirely ignorant
of causality, or the distinction between retarded and advanced Green's
functions.  To get a reasonable answer for $S_2(\omega)$ and other
quantities, we need some recipe for adding the missing information.
Such a recipe is available in the perturbative domain of large
$\omega$, where the integral for $Z_{m,n}(\omega)$ can be evaluated by
stationary-phase approximation around the saddle points of the
integrand.  As in the case of the one-point function, the recipe is to
select those saddle points that contribute with the behavior dictated
by causality.  These are the ``good'' or ``causal'' saddle points, and
they are retained.  The remaining ones are ``bad'' or ``acausal'' and
are thrown away.

Let us review in somewhat more detail how this works. For
concreteness, we fix the sign of the imaginary part of $\omega$ to be
negative.  The saddle-point equation for the integrand $\exp (i\omega
{\rm Tr} \, \Sigma_3 Q /2\alpha)$ is easily seen to be $[Q,\Sigma_3] =
0$.  This equation is solved by the matrix $Q = \Sigma_3$ in all cases
$\beta = 1, 2, 4$.  For the sign choice made, this is a good saddle
point, as is suggested by the observation that the saddle-point value
${\rm e}^{i\omega {\rm Tr}\Sigma_3^2}$ is a {\it maximum} of the
integrand for ${\rm Im}\,\omega < 0$.  (Actually, the true reason is
that the saddle-point value ${\rm e}^{i\omega(m+n)}$ or ${\rm
e}^{2i\omega(m+n)}$ {\it decreases} in magnitude when $\omega$ is
moved into the lower half of the complex plane, for negative $m+n$.
Growth would be incompatible with the analytic properties of the
Green's function.)  By the same token, the saddle point $Q = -
\Sigma_3$ is bad and is thrown away.  When the sign of ${\rm Im}\,
\omega$ is changed, the roles of $Q = \Sigma_3$ and $Q = - \Sigma_3$
are reversed.  Note that the two matrices $Q = \pm \Sigma_3$ are
special in that they are invariant under the action of the symmetry
group $H$, which means they really are isolated saddle {\it points}.

Other solutions $Q_1$, $Q_2$, ... of the saddle-point equation are
obtained by starting from the good $Q_0 \equiv \Sigma_3$ and
exchanging the positions of a small number $p$ (or $2p$) of the $m$
(or 2$m$) entries $+1$ with the same portion of the $n$ (or 2$n$)
entries $-1$.  Since they descend from a causal parent, the saddles so
obtained are expected to be still ``causal'' and are retained.  Their
negatives $-Q_1, -Q_2, ...$ are thrown away.  Unlike $Q_0$, the
diagonal matrices $Q_1, Q_2, ...$ are not stable under the action of 
the symmetry group $H$ but belong to saddle-point {\it manifolds} or
orbits generated by $H$.  Once again, these manifolds belong to the
category of symmetric spaces.  For example, the orbit of $H = {\rm
U}(m) \times {\rm U}(n)$ on $Q_1$ for $\beta = 2$ is
	$$
	\Big( {\rm U}(m) / {\rm U}(m-1) \times {\rm U}(1) \Big)
	\times 
	\Big( {\rm U}(n) / {\rm U}(n-1) \times {\rm U}(1) \Big) \;.
	$$
The contributions of all these saddles to the large-$\omega$
asymptotics of $S_2(\omega)$, or the two-level correlation function
$R_2(\omega)$, have been worked out in Refs. \cite{km1,km2,yl}.  It
was found that only a small number of saddle-point manifolds $(p = 0,
1$ for $\beta = 1, 2$ and, in addition, $p = 2$ for $\beta = 4$)
survive in the limit $m = n = 0$.  The mechanism for termination is
again the dependence on $m$ and $n$ of the (causal) analytic
continuation of the volume of these manifolds.  The asymptotic 
expressions for $S_2(\omega)$ turn out to be
	\begin{eqnarray*}
	&&\beta = 1 : \quad 1-2\,\omega^{-2}+\omega^{-4}{\rm e}^{-2i\omega}
	\\
	&&\beta = 2 : \quad 1-\omega^{-2}+\omega^{-2}{\rm e}^{-2i\omega}
	\\
	&&\beta = 4 : \quad 1-{\textstyle{1\over 2}} \omega^{-2}
	+ (\pi/2) \, \omega^{-1} {\rm e}^{-2i\omega} + (2\omega)^{-4}
	{\rm e}^{-4i\omega} \;.
	\end{eqnarray*}
The real parts of these expressions agree with the asymptotic limits
that are known, for $R_2(\omega)$, from orthogonal polynomials
\cite{mehta} or the supersymmetric me\-thod \cite{efetov}.  It is 
reasonable to expect that this agreement is not accidental but extends
to all orders in the asymptotic expansion in $\omega^{-1}$.  In other
words, we conjecture that the fermionic replica trick in its recent
elaboration is {\it perturbatively equivalent} to the supersymmetric
method.  Also, note that the expression for $\beta = 2$ is not just
asymptotic but is {\it exact}! We will elaborate on this later.

What about the bosonic replica trick?  For the one-point function we
saw that there is a basic asymmetry between bosonic and fermionic
replicas.  The same is true here.  While the use of fermionic replicas
for $Z_{m,n}(\omega)$ leads to a compact symmetric space $X_{m,n}$,
bosonic replicas for $Z_{-m,-n}(\omega)$ lead to a {\it noncompact}
analog space, $Y_{m,n}$ (which is still a Riemannian symmetric space,
albeit of noncompact type.)  The pairing of compact spaces with their
noncompact analogs is given in the following list:
	$$
	\begin{array}{lll}	
	\beta = 1 : ~~
	&{\rm Sp}(2m+2n)/{\rm Sp}(2m) \times {\rm Sp}(2n) ~
	&{\rm O}(2m,2n) / {\rm O}(2m) \times {\rm O}(2n)	
	\\
	\beta = 2 : ~~
	&{\rm U}(m+n) / {\rm U}(m) \times {\rm U}(n) ~
	&{\rm U}(m,n) / {\rm U}(m) \times {\rm U}(n)
	\\
	\beta = 4 : ~~
	&{\rm O}(2m+2n)/{\rm O}(2m) \times {\rm O}(2n) ~
	&{\rm Sp}(2m,2n) / {\rm Sp}(2m) \times {\rm Sp}(2n) \;.
	\end{array}
	$$
The integral representation for $Z_{-m,-n}(\omega)$ in the universal
low-frequency limit is still of the form
	$$
	Z_{-m,-n}(\omega) = \int_{Y_{m,n}} \exp \left( -
	{i\omega \over 2\alpha} {\rm Tr} \, \Sigma_3 Q \right) dQ \;.
	$$
Moreover, the integration manifold $Y_{m,n}$ is still generated by
acting with the noncompact large group $G$ on the diagonal matrix
$\Sigma_3$ by conjugation: $Q = g \Sigma_3 g^{-1}$ $(g \in G)$, and 
we still have the invariant saddle point $Q_0 = \Sigma_3$.  The
large-$\omega$ asymptotic expansion around this saddle point of
$Y_{m,n}$ is known \cite{wegner2} to give the same results as the
expansion around $Q_0 = \Sigma_3$ for $X_{m,n}$, at $m = n = 0$.
What is different is that $Q_0$ now is the {\it only} saddle point.
There are no others that lie directly on $Y_{m,n}$, although there do
exist saddle-point manifolds that can be reached from $Y_{m,n}$ by 
moving a finite distance along an {\it imaginary} direction of the 
real manifold $Y_{m,n}$.  For example, for $\beta = 2$ we set $A =
E_{1,m+1} + E_{m+1,1} \in {\rm Lie} \, {\rm U}(m,n)$ -- where
$E_{p,q}$ is the matrix whose entries are zero everywhere except on
the intersection of the $p$-th row with the $q$-th column where the
entry is unity -- and follow the ``imaginary'' orbit $Q(it) = {\rm
e}^{itA} \Sigma_3 {\rm e}^{-itA}$.  At $t = \pi / 2$ we reach
$Q(i\pi/2) = Q_1$.  By applying the symmetry group $H = {\rm U}(m)
\times {\rm U}(n)$ to $Q_1$, we arrive at an analog of the $p = 1$
saddle-point manifold of the fermionic variant.  If one evaluates
the integrand on this saddle-point manifold, one gets a factor 
${\rm e}^{-i(m+n-2)\omega}$.  For $m = n = 0$, this {\it grows} as
$\omega$ moves into the lower half of the complex plane, which is
unphysical.  Hence, this saddle-point manifold must be discarded, and
we are back to the statement \cite{vz1,vz} that the large-$\omega$
limit of the function $Z_{m,n}(\omega)$ for negative $m$ and $n$ is
exhausted by a single saddle point $(p = 0)$.  In summary, while the
fermionic replica trick suffers from an excess of saddle points, the
bosonic version has ``too few''.

Although the recipe of selecting good saddle points in the fermionic
formulation appears to work, it does need further justification.  A
more proper procedure is to {\it combine} the information from bosonic
and fermionic replicas and write down a bona fide analytic continuation
for $Z_{m,n}(\omega)$ away from the integers.  The proposal made in
\cite{km2,yl} is again to extend the sum over saddle-point manifolds
$p$ all the way up to infinity.  Doing so, however, we run into the
same difficulty we analysed in considerable detail for the one-point
function.  We shall not repeat this analysis here, but only summarize
the facts: for noninteger $m$ and/or $n$, the coefficients in the sum
over $p$ grow in a factorial manner, so that the sum {\it diverges}
and does {\it not} define an analytic continuation of $Z_{m,n}
(\omega)$ to $m,n \notin {\mathbb Z}$.  In the absence of a
well-defined analytic continuation we, of course, have no mathematical
control on the limit $-\lim_{m,n\to 0} (mn)^{-1} \partial^2 Z_{m,n} /
\partial
\omega^2$.  

My interpretation of the divergence is the same as before.  For a
large but fixed value of $\omega$, the saddle-point approximation to
$Z_{m,n} (\omega)$ eventually breaks down (or at least it does for
$\beta = 1, 4$) when $m$ or $n$ becomes too large.  Indeed, for large
$m = n$, we should introduce standard polar coordinates on $X_{m,n}$
to cast $Z_{m,n}(\omega)$ in the form of a Coulomb gas partition
function, and look for a mean field of the Coulomb gas.  Analytic
continuation in the particle number $m = n$ of the Coulomb gas looks
benign.  We therefore expect the divergent sum over saddle-point
manifolds $p$ to get cut off at large values of $m, n$.  The existence
of such a cutoff implies that there exists a nonperturbative
correction which we are missing when writing the sum over saddle-point
manifolds $p$.  For large $\omega$, this correction will not affect
the extrapolation to $m = n = 0$.  On the other hand, for small
$\omega$ the correction may cause an uncontrollable error.

The remaining question is: what happens {\it beyond} perturbation
theory?  Is it reasonable to expect that replicas will penetrate the
nonperturbative small-$\omega$ regime?  This is hard to answer, but
one warning can be issued with certainty.  A series expansion obtained
by saddle-point approximation is almost always {\it asymptotic}, which
means the series {\it diverges}.  What happens is that for a fixed
value of the expansion parameter, $1/\omega$ in our case, the series
initially becomes a better approximation with increasing order, but
eventually turns away and explodes.  Without knowing the exact answer,
it is often difficult to locate the turning point.  For this reason,
asymptotic expansions hardly deserve to be called ``systematic''.  The
only safe way of using an asymptotic series is to {\it fix} the order
of approximation and then lower the expansion parameter $1/\omega$
accordingly, so as to make the error term negligible.  This limits the
usefulness of asymptotic series in practice.  To do better, one needs
to establish Borel summability and carry out Borel resummation.  This
is already nontrivial for integrals with one saddle point, and becomes
much more difficult when two or more saddle points are involved.  On
the other hand, not all saddle-point approximations are asymptotic, as
is demonstrated by the case of the two-point function for $\beta = 2$.
We will explain in the next section what is special about that
example.

The nonperturbative ambiguities of the replica trick for disordered
electron systems are not restricted to the zero-dimensional limit
but also affect the replica field theory for $d$-dimensional systems.
Let me finish the section with an example demonstrating this.

Disordered two-dimensional electrons in a strong magnetic field
display the (integer) quantum Hall effect.  A replica field theory for
such systems was proposed long ago by Pruisken \cite{pruisken}.  The
theory is a compact nonlinear $\sigma$ model with Lagrangian
	$$
	L = {\sigma_{xx} \over 8} {\rm Tr} \, \partial_\mu Q \,
	\partial_\mu Q + {\sigma_{xy} \over 8} \epsilon^{\mu\nu}
	{\rm Tr}\, Q \, \partial_\mu Q \, \partial_\nu Q \;,	
	$$
where the field $Q = U \Sigma_3 U^{-1}$ parametrizes ${\rm U}(m+n)
/ {\rm U}(m) \times {\rm U}(n)$.  Without loss in the zero replica
limit, we may take $m = n$.  The coupling constants $\sigma_{xx}$ and
$\sigma_{xy}$ are interpreted as the conductivities (dissipative and
Hall) of the noninteracting electron gas. The Hall conductivity
$\sigma_{xy}$ determines the current response transverse to an applied
electric field.  In particular, it determines the ${\it orientation}$
of the current flow around the boundary of a finite sample \cite{xrs}.

Now comes a disaster.  Use of the isomorphism $Q \mapsto -Q$ in
Pruisken's Lagrangian changes the sign of the term $\epsilon^{\mu\nu}
{\rm Tr} \, Q \partial_\mu Q \partial_\nu Q$.  Thus the field theory
coupling $\sigma_{xy}$ is defined only up to a sign.  For $m=n=1$,
where the above field theory (with $\sigma_{xy} = S$) is known 
\cite{haldane} to describe antiferromagnetic quantum spin-$S$ chains, 
this indeterminacy makes sense.  Antiferromagnets carry no sense of
orientation.  But for $n = 0$, $\sigma_{xy}$ is supposed to be the
physical Hall conductivity, so the field theory makes the nonsensical
prediction that the orientation of the current response of a quantum
Hall sample is ill-determined?!

Of course, this is just another manifestation of the causality problem
we have been emphasizing all along.  Can it be cured?  Pruisken's
field theory is known to flow to the strong coupling (or small
$\sigma_{xx}$) regime, where the dominant field configurations form a
dense gas of interacting instantons and anti-instantons.  In such a
nonperturbative soup of saddle points and almost-saddle points, it
seems quite hopeless to try and disentangle the good configurations
from the bad ones.

\section{Semiclassical exactness for $\beta = 2$}
\label{sec:DH}

We have seen that the fermionic replica trick, augmented by an
inspired recipe for selecting good saddle points, for $\beta = 2$
already gives the correct answer for the universal limit of
$S_2(\omega)$ when the saddle-point approximation is carried out to
leading order in the small parameter $1/\omega$.  Apparently, the
saddle-point approximation in this case is exact, not just
approximate! We call integrals where this miracle happens
``semiclassically exact''. In the present section we will explain the
mathematical basis underlying the phenomenon of semiclassical
exactness: the Duistermaat-Heckman theorem.  This is a celebrated
result in symplectic geometry \cite{dh,witten1,ab,vergne} and is
included here for the convenience of the targeted reader.  A few
references congenial to physicists are \cite{stone,bt,szabo}.

Let $M$ be a symplectic manifold of dimension $2f$.  In physics such a
manifold is called a phase space with $f$ degrees of freedom.  Simple
examples are the two-sphere ${\rm S}^2$ or the two-torus ${\rm T}^2$.
The example we are particularly concerned with is the coset space
${\rm U}(m+n) / {\rm U}(m) \times {\rm U}(n)$, which has dimension $2f
= 2mn$.  A symplectic manifold $M$ comes with a symplectic structure,
{\it i.e.},  a closed and nondegenerate two-form $\Omega$.  For the
two-sphere, $\Omega$ is the solid angle, expressed in spherical polar
coordinates $\theta, \phi$ by $\sin\theta \, {\rm d}\theta \wedge {\rm
d}\phi$.  For the space ${\rm U}(m+n) / {\rm U}(m) \times {\rm U}(n)$
parametrized by $Q$, the symplectic structure is $(8i)^{-1} {\rm Tr}
\, Q {\rm d}Q \wedge {\rm d}Q$.  By Darboux's theorem, there exist 
local coordinate systems made up of positions $q^i$ and momenta $p_i$,
such that $\Omega$ takes the canonical form $\Omega = \sum_{i=1}^f
{\rm d}p_i \wedge {\rm d}q^i$.  Let $\Omega^f = \prod_{i=1}^f {\rm
d}p_i \wedge {\rm d}q^i$ denote the Liouville form on $M$.  In the
case of interest, $\Omega^f$ agrees with the volume element $dQ$ up to
a multiplicative constant.

The Duistermaat-Heckman theorem in the formulation given below makes a
statement about integrals of the form
	$$
	Z(\omega) = \int_M \Omega^f {\rm e}^{i\omega H} \;,
	$$
where the function $H : M \to {\mathbb R}$ is required to possess the
following property.  Viewing $H$ as a Hamiltonian function, we have
the canonical equations of motion of Hamilton mechanics:
	$$
	\dot q^i = {\partial H \over \partial p_i} \;, \quad
	\dot p_i = - {\partial H \over \partial q^i} \;.
	$$
The solutions of these equations are conveniently assembled into a map
$M \times {\mathbb R} \to M$, $(x,t) \mapsto \psi_t(x)$, called the
phase flow of $H$.  The crucial property we shall require of $H$ is
that there exist some {\it Riemannian metric} on $M$ which is preserved
by the flow $\psi_t$.

Before stating the Duistermaat-Heckman theorem, let us look at a few
examples where this condition is satisfied.  The simplest one is
provided by the Hamiltonian for a classical spin in a magnetic field
\cite{stone}.  The phase space $M$ in this case is the two-sphere
${\rm S}^2$, and the motion is precession of the spin around the axis
of the magnetic field.  The frequency of precession, the Larmor
frequency, does not depend on the tilt angle between the magnetic
field axis and the axis of the spin.  Thus the phase flow $\psi_t$ is
simply uniform rotation around the field axis, with the rotation angle
being a linear function of time.  Clearly, this flow preserves the
natural metric ${\rm d}\theta^2 + \sin^2 \theta \, {\rm d}\phi^2$ of
${\rm S}^2$.

The classical spin in a magnetic field is but one of a large family of
integrable Hamiltonian systems with the same property.  A brief sketch
is as follows.  Let $G$ be a compact semisimple Lie group with maximal
abelian subgroup (or maximal torus) $T$.  We fix some regular element
$A$ of the Lie algebra ${\rm Lie}(T)$, and consider the adjoint orbit
${\rm Ad}(G) A$ consisting of elements $UAU^{-1}$.  The adjoint orbit
comes with a natural Hermitian structure, {\it i.e.}, a symplectic
structure $\Omega$ as well as a metric tensor $g$, which we refrain
from writing down here as this would lead us just a little too far
(for these details consult Appendix 2 of \cite{arnold}).  Note that we
get the phase space ${\rm S}^2$ of the classical spin by putting $G =
{\rm SU}(2)$ and $A = \sigma_3$.  Now we pick a quadratic form, say
$\langle X , Y \rangle = {\rm Tr} \, XY$, on the Lie algebra ${\rm
Lie}(G)$, and we fix another regular element $B$ of ${\rm Lie}(T)$ to
form the function $H = {\rm Tr} \, B U A U^{-1}$.  We may view the
adjoint orbit ${\rm Ad}(G) A$ as a phase space, with $H$ being a
Hamiltonian function on it.  The phase flow of this Hamiltonian turns
out to be $\psi_t : U A U^{-1} \mapsto {\rm e}^{tB} U A U^{-1} {\rm
e}^{-tB}$, and this flow preserves the Riemannian metric $g$ of the
adjoint orbit.  We are mentioning this class of examples because it
will be seen to give rise to the Itzykson-Zuber formula.

Our example of interest is the function $H = {\rm Tr} \, hQ/2$, with 
	$$ h = {\rm diag}(h_+,h_-) = 
	{\rm diag}(h_{+1},...,h_{+m};h_{-1},...,h_{-n}) 
	$$ 
a diagonal real matrix, on the symplectic manifold ${\rm U}(m+n) / {\rm
U}(m) \times {\rm U}(n)$.  (This example can actually be regarded as a
degenerate limit of the above general family.) To construct its phase
flow in an elementary fashion, we express $Q$ as
	$$ 
	Q = \pmatrix{ 1 - 2BB^\dagger & 2B\sqrt{1 - B^\dagger B} 
	\cr 2B^\dagger \sqrt{1 - BB^\dagger} &-1+2B^\dagger B\cr} \;, 
	$$ 
where $B$ is a complex $m \times n$ matrix with range $0 \le B^\dagger
B \le 1$. The symplectic structure and the Hamiltonian for this choice
of coordinates take the form
	\begin{eqnarray*} 
	\Omega &=& (8i)^{-1} {\rm Tr} \, Q {\rm d}Q \wedge {\rm d}Q 
	= i^{-1} {\rm Tr} \, {\rm d}B \wedge {\rm d}B^\dagger \;, \\ 
	H &=& {\rm Tr} \, hQ/2 = {\rm const} - {\rm Tr} \, h_+ BB^\dagger 
	+ {\rm Tr} \, h_- B^\dagger B \;.  
	\end{eqnarray*} 
The phase flow $\psi_t$ is then readily seen to be $B \mapsto {\rm
e}^{ith_+} B {\rm e}^{-ith_-}$, or in the original representation, 
	$$
	\psi_t : Q \mapsto {\rm e}^{ith} Q {\rm e}^{-ith} \;.  
	$$
Clearly, this preserves the natural Riemannian metric ${\rm Tr} \,
{\rm d}Q^2$.  

We now come to the Duistermaat-Heckman theorem.  The theorem is rooted
in equivariant cohomology, and its best version relies on no more than
Stokes' formula for differential forms that are equivariantly closed
for a compact group action.  In particular, no reference to any
metric structure is needed.  Nevertheless, to avoid mathematical
overload, we shall permit ourselves the luxury of using a metric.

{\bf Theorem.}  Let $(M,\Omega,H)$ be a Hamiltonian system with
compact $2f$-dimen\-sional phase space $M$, and let its phase flow
preserve a Riemannian metric on $M$.  Then the integral $\int_M
\Omega^f {\rm e}^{iH}$ localizes on the critical set of $H$.

{\bf Remark.}  What the theorem is saying is that the integral is
completely determined by the values of the function $H$, and a finite
number of derivatives thereof, on the set of solutions of ${\rm d}H =
0$ (the saddle points or saddle-point manifolds).  In other words, the
stationary-phase approximation is exact.

{\bf Proof.}  We will employ a method of proof originally due to Bismut
\cite{bismut} (see also Refs. \cite{witten2,sz}), which uncovers and 
takes advantage of a hidden supersymmetry.  Let $x^i$ $(i=1,...,2f)$
be a system of local coordinates of $M$, in which the symplectic
structure is expressed by $\Omega = {1\over 2} \Omega_{ij} (x) \, {\rm
d}x^i\wedge {\rm d}x^j$ (summation convention).  We supersymmetrically
extend the phase space $M$ by introducing for every $x^i$ a
corresponding anticommuting coordinate $\xi^i$.  Consider then the
expression
	$$ 
	\int \exp \left( - {1\over 2} 
	\Omega_{jk}(x) \xi^j \xi^k + i H(x) \right) 
	$$ 
where $\int$ means integration with the flat Berezin form ${\rm d}
x^1\ldots {\rm d}x^{2f} \partial_{\xi^1} \ldots \partial_{\xi^{2f}}$.
By using canonical coordinates, it is not difficult to see that the
Fermi integral, {\it i.e.},  differentiation with respect to the
anticommuting variables and multiplication by ${\rm d}x^1 \ldots {\rm
d} x^{2f}$, produces the Liouville form $\Omega^f$.  Hence we have
	$$
	\int \exp \left( - {1\over 2} \Omega_{jk} \xi^j \xi^k
	+ i H \right) = \int_M \Omega^f {\rm e}^{iH} \;.
	$$
Next, using the closedness of $\Omega$ ($\xi^j \xi^k \xi^l \partial
\Omega_{jk} / \partial x^l = 0$), one verifies that the exponent 
$S \equiv -{1 \over 2} \Omega_{jk} \xi^j \xi^k + iH$ has the property
of being annihilated by
	$$ 
	D = \xi^j {\partial\over\partial x^j} + i \Omega^{jk}
	{\partial H \over \partial x^j} {\partial\over\partial\xi^k}
	\;. 
	$$ 
$D$ is a first-order differential operator of odd type, and is 
analogous to the BRST operator that plays a central role in the 
functional integral quantization of nonabelian gauge theories 
\cite{zinnjustin}.  The key step now is to deform the integral by a 
parameter $t$:
	$$
	\int_M \Omega^f {\rm e}^{iH} = \int {\rm e}^S
	= \int {\rm e}^{S + tD\lambda} \;,
	$$
where the function $\lambda$ is constrained by $D^2 \lambda = 0$ and 
will be specified presently.  The Berezin integral on the right-hand side 
has the crucial property of being {\it independent} of $t$.  Indeed, by 
Taylor expanding,
	$$
	\int {\rm e}^{S + tD\lambda} = \int {\rm e}^S
	\Big( 1 + tD\lambda + \textstyle{1\over 2} t^2 (D\lambda)^2 
	+ \ldots \Big) \;,
	$$
and using $D^2 \lambda = 0$ and partial integration in conjunction 
with $DS = 0$, one sees that the terms of linear and higher order in 
$t$ all vanish.  To reap full benefit from the $t$-independence of
the integral, we pick a well-chosen Riemannian metric $g$ and set
	$$
	\lambda = -i g^{jk} \Omega_{kl} 
	{\partial H \over \partial x^j} \xi^l \;.
	$$ 
It can be shown that this expression satisfies the condition $D^2
\lambda = 0$ if and only if the metric $g$ is invariant under the
phase flow of $H$.  By assumption, such a metric $g$ exists. The last 
step is to take the parameter $t$
to infinity.  Since the number part of
	$$
	D\lambda = - g^{jk} {\partial H \over \partial x^j}
	{\partial H \over \partial x^k} + {\cal O}(\xi\xi)
	= - |{\rm d}H|^2 + {\cal O}(\xi\xi)
	$$
is negative definite, this limit localizes the integral $\int_M
\Omega^f {\rm e}^{iH} = \int {\rm e}^{S + tD\lambda}$ onto the 
critical set of $H$, {\it i.e.}, the solutions of the saddle-point
equation ${\rm d}H = 0$.  This concludes the proof.

If the saddle points of $H$ are isolated, one can easily push the
calculation further and write down an explicit formula for the 
integral as a discrete sum:
	$$
	\int_M \Omega^f {\rm e}^{iH} = 
	(2\pi)^f \sum_{x : {\rm d}_x H = 0} 
	i^{{1 \over 2}{\rm sgn}({\rm Hess}_x H)}
	{{\rm e}^{iH(x)} \over \sqrt{ | {\rm Det}({\rm Hess}_x
	H) |}} \;,
	$$
where the phase of the contribution from each saddle point is
determined by the signature of the Hessian of $H$ at $x$.  For
the classical spin in a magnetic field, this gives
	$$
	\int\limits_{{\bf n}^2 = 1} {\rm e}^{i {\bf n} \cdot
	{\bf B}} d^2 n = 2\pi \, {{\rm e}^{i|B|} - {\rm e}^{-i|B|}
	\over i|B|} \;,
	$$
the correctness of which is easily verified by direct computation.
Less trivial integral formulas result on taking for the Hamiltonian
system $(M,\Omega,H)$ the adjoint orbit of a compact Lie group $G$,
with the Hamiltonian being $H = {\rm Tr} \, A U B U^{-1}$.  If both
$A$ and $B$ are regular elements in a Cartan subalgebra of $G$, 
one obtains \cite{dh,vergne}
	$$
	\int_G {\rm e}^{{\rm Tr} \, A U B U^{-1}} dU = 
	\big( p(A) p(B) \big)^{-1} \sum_{w\in W_G} (-1)^{|w|}
	{\rm e}^{{\rm Tr} \, A \, wB} \;,
	$$
where $p(A) = \prod_{\alpha > 0} \alpha(A)$ is a product over positive
roots, the sum runs over the Weyl group \cite{helgason} of $G$, and
$|w|$ denotes the parity of $w$.  (Regularity of $A$ means that $p(A)$
is nonzero.)  $dU$ is a suitably normalized Haar measure of $G$.  Note
that the integral can and should actually be understood as being over 
the symplectic quotient $G/T$.  That the factor of $i$ has disappeared
from the exponent is of no concern. Looking back at the proof of the
Duistermaat-Heckmann theorem, we see that $i$ didn't play any role:
the underlying principle is localization on the critical set, not
stationary phase.

The above formula is due to Harish-Chandra (1957) who proved it by a
quite different method, namely by computing the radial parts of
$G$-invariant differential operators \cite{hc}.  Specializing to the
case $G = {\rm U}(N)$ (or ${\rm SU}(N)$, it doesn't matter) and
setting $A = {\rm diag}(A_1, \ldots , A_N)$, $B = {\rm diag}(B_1,
\ldots , B_N)$ we get
	$$
	\int_{{\rm U}(N)} {\rm e}^{{\rm Tr} \, A U B U^{-1}} dU =
	{{\rm Det}({\rm e}^{A_i B_j})_{i,j=1,...N} \over
	\prod_{i < j} (A_i - A_j) (B_i - B_j)} \;,
	$$
which is known in physics as the Itzykson-Zuber formula.

We finally understand the semiclassical exactness of the replica
integral for the $\beta = 2$ two-point function in the universal
limit.  The nonlinear $\sigma$ model manifold ${\rm U}(m+n) / {\rm
U}(m) \times {\rm U}(n)$ is a $2mn$-dimensional phase with symplectic
structure $(8i)^{-1} {\rm Tr} \, Q {\rm d}Q \wedge {\rm d}Q$, and the
flow of the ``Hamiltonian'' $H = \omega {\rm Tr} \, \Sigma_3 Q / 2$
preserves a Riemannian metric $g = {\rm Tr} \, {\rm d}Q^2$.
Therefore, the Duistermaat-Heckman localization principle applies, and
the integral $\int \exp ( i\omega {\rm Tr} \, \Sigma_3 Q / 2) \, dQ$
is evaluated exactly by saddle-point approximation.

\section{Conclusion}
\label{sec:summary}

The replica trick for spectral correlations of disordered electron
systems is mathematically ill-founded because analytic continuation of
the generating function
	$$
	Z_{m,n}(u,v) = \Big\langle {\rm Det}^m (u-H)
	\, {\rm Det}^n (v-H) \Big\rangle
	$$
away from the integers $m,n$ is not unique.  In benign cases, such as
the famous Selberg integral, uniqueness is guaranteed by a boundedness
property as spelled out by Carlson's theorem or Carleman's theorem 
\cite{titchmarsh}.  It appears that no such property is available in 
the present case. (However, it has been suggested to me \cite{jac} that 
this problem might be overcome by considering a generating function of 
the positive real form
	$$
	\Big\langle {\rm Det}^m \big((E_1-H)^2 + \epsilon^2 \big)
	\, {\rm Det}^n \big( (E_2-H)^2 + \epsilon^2 \big) \Big\rangle \;.)
	$$

For positive $m,n$ (the fermionic replica trick) the generating
function is {\it analytic} in $u$ and $v$, which means that the entire
information about causality of Green's functions is missing.  (Of
course, the enlightened user knows about causality, but fermionic
replicas by themselves do not.)  In recent work by Kamenev and Mezard,
and by Yurkevich and Lerner, causality was introduced into the
formalism by selecting a well-chosen set of saddle points of the
compact nonlinear $\sigma$ model based on fermionic replicas.  In this
way, the known large-frequency asymptotics of the level correlation
functions for disordered electron systems was reproduced for all
cases $\beta = 1, 2, 4$.  This motivates the conjecture that the
fermionic replica trick (augmented by the selection of ``causal''
saddle points) is {\it perturbatively equivalent} to the
supersymmetric method.

Close inspection of Refs. \cite{km1,km2,yl} reveals that the procedure
used there is mathematically uncontrolled, as it involves a
nonexistent analytic continuation.  The fact of the matter is that the
sum over saddle-point manifolds $p = 0, 1, ... \, \infty$ {\it
diverges} for noninteger $m, n$.  We have argued that this divergence
is likely to be cut off by a nonperturbative mechanism.  Although such
a cutoff is immaterial for the analytic continuation to $m = n = 0$
for large frequency $\omega$, it may introduce uncontrollable errors
for small $\omega$.  In any case, the method as it stands is limited
to the perturbative regime, as it relies on stationary-phase
evaluation of integrals, which requires a small parameter $1/\omega$.

A miraculous exception is Wigner-Dyson statistics for $\beta = 2$,
where a hidden supersymmetry, namely equivariant cohomology and the
localization principle underlying the Duistermaat-Heckman theorem,
localizes the nonlinear $\sigma$ model integrals on the saddle points.
Put differently, the leading-order stationary-phase approximation in
this case is {\it exact}, regardless of whether $\omega$ is large or
small.  The same mechanism is at work in the supersymmetric
formulation.

\end{document}